\documentclass[seceq]{ptptex}
\title{%        %You can use \\ for explicit line-break
Coherent Tunneling and Instantons in Presence of Classical Chaos %
}

\author{%       %Use \scshape  for the family name
Viatcheslav \textsc{Kuvshinov}$^1$ and Andrei \textsc{Kuzmin}$^2$%
}

\inst{%         %Affiliation, neglected when [addenda] or [errata]
$1$-Institute of Physics, 220072 Belarus, Minsk, Scorina,68.
                    Tel. 375-172-84-16-28, fax: 375-172-84-08-79.
                    E-mail: kuvshino@dragon.bas-net.by

$2$-Institute of Physics, 220072 Belarus, Minsk, Scorina,68.
                    Tel. 375-172-84-16-28.
                    E-mail: avkuzmin@dragon.bas-net.by }

\abst{%         %this abstract is neglected when [addenda] or [errata]
We consider the instanton approach to the problem of chaos assisted tunneling in the
context of existing analytical and numerical results obtained in this field. We
provide the estimation for the range of validity of this method and briefly discuss
possible applications of chaos assisted tunneling for quantum computations. }

\begin{document}

\maketitle

\section{Introduction}

The influence of classical chaos on dynamical tunneling~\cite{Heller} in the
semiclassical regime of quantum mechanics was observed in the pioneer work by Lin and
Ballentine~\cite{Lin}. Namely, it was detected in the numerical experiment that
regions of chaotic motion in the classical phase space of the model system
(periodically driven quartic double well) can enhance the rate of quantum tunneling
between KAM-islands in several orders of magnitude compared to the ordinary tunneling
in completely integrable case. The coherent nature of the tunneling was also
emphasized (no leakage of the probability into the chaotic zone). The opposite impact
of classical chaos on dynamical tunneling was found in the Ref.~\cite{Grossman}. On
the example of the model system with double-well potential affected by the periodic in
time perturbation it was numerically demonstrated that for specific parameter values
the tunneling between symmetry related KAM-islands can be coherently destructed i.e.
the wave packet can be almost completely localized in one of the wells. For the
theoretical explanation of the chaos assisted tunneling the multi-level
model~\footnote{The three level-model is mainly exploited, since the singlet-doublet
crossings are more likely to occur in the (quasi-)energy spectrum.} was proposed and
tested for the two coupled quartic oscillators~\cite{Bohigas}. The explanation of the
tunneling rate enhancement observed in the numerical simulations is the replacement of
usual two-level tunneling dynamics by more complicated three-level tunneling
mechanism, appeared at the avoided level crossing of primarily regular (quasi-)energy
doublet and chaotic eigenstate~\cite{Latka1}. Detailed description of numerical
technique which can be used for simulation of chaos assisted tunneling can be found in
the Ref.~\cite{Utermann}. Possible applications of the chaos assisted tunneling are
the active product selection in chemical reactions and the dynamical tunneling control
for various physical systems~\cite{Latka2}.

The Refs.~\cite{Lin,Grossman,Bohigas,Latka1,Utermann,Latka2} are clearly indicate the
existing priority of numerical simulations for the investigation of the chaos assisted
tunneling regime. A very few works devoted to the analytical investigation of this
phenomenon do exist. Namely, the path integral formalism was primarily used for the
analytical description of the dynamical tunneling between symmetry related KAM-tori in
the annular billiard possessing the mixed dynamics in the Ref.~\cite{Doron}.
Description of chaotic tunneling in terms of classical chaotic manifolds in the
complex regime (complexified phase space) was done in the Ref.~\cite{Shudo}. Instanton
technique~\cite{BPST}, primarily devised for quantum gauge field theories, can be used
for more deep understanding of the chaos assisted tunneling regime in terms of path
integral formulation of quantum mechanics. Particulary, chaotic instanton solutions,
primarily obtained in the Ref.~\cite{Acta}, seem to play important role in dynamical
tunneling through chaotic layer~\cite{PRE}.

The development of new analytical approaches (see also Refs~\cite{Aoki,Rubin}) is
motivated by the following facts. Numerical simulations in the main are based on the
Floquet theory~\cite{Shirley}. One-dimensional systems affected by the monochromatic
time-dependent perturbation (the aliquot frequencies do not change the picture) were
mainly considered~\cite{Lin,Grossman,Bohigas,Latka1,Utermann,Latka2}. While the number
of independent frequencies in the Fourier spectrum grows (or the number of degrees of
freedom of the system increases) the size of the Floquet Hamiltonian matrix grows
under the exponential law. Therefore, more computer resources are needed. In this case
analytical methods including instanton approach may play an important role.

Another reason is based on the fact that numerical calculations can not give the
complete picture, because it is impossible to cover in numerical simulations the whole
continuous parametric region, the common features between different classes of systems
can also escape one's attention. It does not underestimate the importance of numerical
studies but rather states the necessity of analytical ones.

\section{Instanton approach to chaos assisted tunneling}

The instanton method showed oneself well in description of the spectral properties and
tunneling probabilities between lower eigenstates~\cite{Rajar}. The attempt to
generalize the instanton method for non-autonomous Hamilton systems was made in the
Refs.~\cite{Acta,PRE}. In the present work we estimate the parametric region where
calculations made in Refs.~\cite{Acta,PRE} are valid. For distinctness we consider the
system with the Euclidean Hamiltonian $H = p^{2}/2 - \omega_{0}^{2}\cos{x} + \epsilon
x \cos{\nu t}$. Here the coordinate $x$ varies from $-\infty$ to $+\infty$ and $t$ is
real and denotes Euclidian time. This Hamiltonian differs from one used in
Refs.~\cite{Acta,PRE} by the form of time-dependent perturbation. The results obtained
in these papers remain valid for this case up to the numerical factor. It emphasizes
their applicability for a wide class of the systems with periodic in space potentials
(single potential well in each period) affected by the small periodic in time
perturbation. The reason is the universality of the separatrix destruction
mechanism~\cite{Sagdeev}. Limitations to the range of validity of the particular
realization of the instanton method and the results obtained in Refs.~\cite{Acta,PRE}
come from the following assumptions made. Stochastic layer is narrow and homogeneous,
therefore the stability island near the elliptic point of the first resonance has to
be indistinguishable. Euclidian actions of the perturbed (chaotic) single-instanton
configurations approximately equal the Euclidian actions of non-perturbed instantons
at non-minimal energies (see Ref.~\cite{PRE}). These conditions lead to the
restrictions~\cite{Sagdeev}:
\begin{equation}\label{cond1}
\overline{\epsilon} \ll \alpha \ll 1, \quad \nu > \omega_0 , \quad \alpha =
\frac{I}{\omega(I)} \left| \frac{d \omega }{dI} \right|.
\end{equation}
Here $\overline{\epsilon} \equiv \epsilon / \omega^2_0$ denotes the dimensionless
coupling constant, and $\alpha$ is the parameter of nonlinearity~\cite{Sagdeev}, $I$
is the action variable, $\omega(I)$ is the frequency of nonlinear oscillations.

For indistinguishability of the first resonance additional restrictions are needed.
The first resonance width in frequency estimated by means of the standard
technique~\cite{Sagdeev} is given by the expression:
\begin{equation}
\delta \omega \sim (\epsilon \omega^{\prime}_1)^{\frac{1}{2}}, \quad \omega^{\prime}_1
\equiv \frac{d \omega (I)}{dI}\left|_{I=I_1} \right. .
\end{equation}
The value $I_1$ is determined by the resonance condition $\omega (I_1) = \nu$.
Therefore the first resonance width in action is estimated as $\delta I = (\epsilon/
\omega^{\prime}_1)^{\frac{1}{2}}$. The condition for indistinguishability of the first
resonance is $\delta I \ll \omega_0$. Thus the system independent restriction has the
form:
\begin{equation}\label{cond2}
 \overline{\epsilon}^{\frac{1}{2}} \ll 1.
\end{equation}
The conditions~(\ref{cond1}, \ref{cond2}) define the parametric range for the
instanton technique used in the Refs.~\cite{Acta,PRE} to be valid.

Let us discuss the prospects for chaos assisted tunneling to be applied for the
enhancement of the stability of quantum computations and reliability of quantum
computers.  Quantum computer consists of qubits (quantum bits) which states
superposition encodes the initial and final data. Quantum computation operations are
realized by the unitary operators (consisted of the elementary operations -- quantum
gates) acting on these superimposed states. There are two main obstacles for
implementation of quantum computations in practice. One of them is the problem of
decoherence~\cite{Decoh}, which means the superimposed state to be destructed rapidly
due to unavoidable interaction with the environment, and to be converted into the
mixture of the qubit states. It destructs the quantum computation process. Another
problem is the existence of small unknown or uncontrollable residual interaction among
qubits~\cite{Prozen1}. Assume that the decoherence process is slow enough (because of
some reasons which are not discussed here) and does not break down quantum
computations. Unitary evolution operator has the form $U = Exp(i H)$, where $H$ is the
Hamiltonian, which is known if one wishes to receive the definite result. However,
residual interactions between qubits lead the Hamiltonian to be consisted of two parts
$H = H_0 + V$. The first part $H_0$ is "large" in the sense it is mainly determine
quantum evolution. The second part $V$ plays the role of small unknown perturbation
originated from the uncontrollable interaction between qubits. For reliability of
results obtained on quantum computer quantum calculations have to be stable with
respect to these perturbations. Quantitative parameter called fidelity determining the
stability of computational process was introduced and widely exploited~\cite{Prozen2}.
The residual interaction between qubits can originate from the tunneling transitions
between them. Thus it is important to suppress such undesirable tunneling coupling. It
can be achieved in chaos assisted tunneling regime~\cite{Grossman}. Therefore the
investigation of the destructive interference of chaotic instanton contributions is of
importance.

\section{Conclusion}

We have demonstrated the necessity of analytical approaches to the tunneling problem.
Application of the instanton method to the problem of chaos assisted tunneling was
discussed in the context of existing results. Chaotic instanton
solutions~\cite{Acta,PRE} may be useful particulary in investigation of chaos assisted
tunneling in one-dimensional systems with degenerate classical ground state affected
by non-monochromatic time-dependent perturbation (having two or more independent
frequencies in its Fourier spectrum). Consideration of such systems seems attractive,
since it was demonstrated that the control of dynamical tunneling can be achieved in
bichromatically driven pendulum~\cite{Latka2}.

We provided the estimation for the range of validity of perturbed instanton approach
for a wide class of one-dimensional systems with periodic potential affected by
periodic in time perturbation. Corresponding conditions for the system parameters were
formulated. Nevertheless for further justification and development of the method the
comparison with numerical data and existing analytical methods is needed. Possible
application of chaos assisted tunneling in connection with quantum computer
stabilization was also briefly discussed.

\section*{Acknowledgements}

The grant of the World Federation of Scientists is gratefully acknowledged.

%\section*{Acknowledgements}
%We would like to thank ...........

%\appendix
%\section{First Appendix} %Empty argument \section{} yields `Appendix'.
%
%\section{Second Appendix}

\end{document}